\documentclass[aps,pra,reprint,superscriptaddress,floatfix]{revtex4-2}

\usepackage[colorlinks=true,urlcolor=blue]{hyperref}
\usepackage{amsmath,amssymb,bm}
\usepackage{graphicx}
\usepackage{float}
\usepackage{cancel}
\usepackage{physics}

\begin{document}

\title{Crossing the superfluid-supersolid transition of an elongated dipolar condensate}
\author{Aitor Alaña}
\affiliation{Department of Physics, University of the Basque Country UPV/EHU, 48080 Bilbao, Spain}
\affiliation{EHU Quantum Center, University of the Basque Country UPV/EHU, Leioa, Biscay, Spain}
\author{Nicolò Antolini}
\affiliation{CNR-INO, Sede di Pisa, 56124 Pisa, Italy}
\affiliation{LENS, University of Florence, 50019 Sesto Fiorentino, Italy}
\author{Giulio Biagioni}
\affiliation{CNR-INO, Sede di Pisa, 56124 Pisa, Italy}
\affiliation{Department of Physics and Astronomy, University of Florence, 50019 Sesto Fiorentino, Italy}
\author{Iñigo L. Egusquiza}
\affiliation{Department of Physics, University of the Basque Country UPV/EHU, 48080 Bilbao, Spain}
\affiliation{EHU Quantum Center, University of the Basque Country UPV/EHU, Leioa, Biscay, Spain}
\author{Michele Modugno}
\affiliation{Department of Physics, University of the Basque Country UPV/EHU, 48080 Bilbao, Spain}
\affiliation{EHU Quantum Center, University of the Basque Country UPV/EHU, Leioa, Biscay, Spain}
\affiliation{IKERBASQUE, Basque Foundation for Science, 48013 Bilbao, Spain}

\date{\today}

\begin{abstract}
We provide a theoretical characterization of the dynamical crossing of the superfluid-supersolid phase transition for a dipolar condensate confined in an elongated trap, as observed in the recent experiment by G. Biagioni \textit{et al.} [Phys. Rev. X \textbf{12}, 021019 (2022)]. By means of the extended Gross-Pitaevskii theory, which includes the Lee-Huang-Yang quantum fluctuation correction, we first analyze the ground state configurations of the system as a function of the interparticle scattering length, for both trap configurations employed in the experiment. Then, we discuss the effects of the ramp velocity, by which the scattering length is tuned across the transition, on the collective excitations of the system in both the superfluid and supersolid phases.
We find that, when the transverse confinement is sufficiently strong and the transition has a smooth (continuous) character, the system essentially displays a (quasi) 1D behavior,  its excitation dynamics being dominated by the axial breathing modes. Instead, for shallower transverse trapping, when the transition becomes discontinuous, the collective excitations of the supersolid display a coupling with the transverse modes, signalling the onset of a dimensional crossover.
\end{abstract}

\maketitle

\section{Introduction}

Supersolids are an exotic phase of matter combining superfluid properties (phase coherence and frictionless flow \cite{Gross62,pitaevskii2016}) with the translational symmetry breaking that characterizes crystalline structures \cite{Gross57,cristSov70,boni12}. 
Firstly predicted in the 50s \cite{Le70,chester70}, supersolids have gained the interest of the scientific community as a consequence of their recent experimental realizations in dipolar condensates
\cite{Ta19,bottcher2019,Ch19,Ta19b,Gu19,Na19,Ta21,No21,sohmen21} and in other ultracold atomic systems \cite{leo17,Li17,schuster20}.  
In particular, Ref. \cite{biagioni2022}  has  recently reported the experimental investigation of the superfluid-supersolid quantum phase transition in an elongated dipolar condensate, driven by tuning the interparticle interactions (by means of Feshbach resonances). 
Remarkably, it has been shown that the character of the transition can be changed from continuous to discontinuous simply by tuning the transverse confinement (or the atom number), therefore providing a dimensional crossover between second-order transitions in 1D \cite{Se08,Pe21} and first-order transitions in 2D \cite{Po94,Ma13,Lu15,Ta19,Bo19,Ta19b}. 

In the present work we provide a complementary theoretical characterization of the equilibrium and dynamical properties of the dipolar condensate in the two trap configurations, $V_C$ and $V_D$, employed in the experiment of Ref. \cite{biagioni2022}. 
This analysis is carried out within the standard framework of the extended Gross-Pitaevskii theory \cite{dalfovo99}, including both contact and dipolar interactions \cite{Ronen:2006}, as well as the Lee-Huang-Yang quantum correction \cite{Wachtler:2016}.  
We first consider the equilibrium properties of this system and we show that the trap characterized by a tighter transverse confinement ($V_C$) presents a smooth/continuous transition between the superfluid and supersolid phases, whereas for shallower trapping potentials ($V_D$) a discontinuous character clearly shows up. Then, we thoroughly discuss how the velocity of the ramp employed to experimentally tune the $s-$wave scattering length across the transition affects the dynamical response of the system, commenting also on the role of the formation time of the supersolid. Remarkably, we find that -- when the system enters the supersolid phase -- the collective modes in the two traps present distinctive behaviors: in the trap $V_C$, where the system can be considered effectively quasi 1D, the excitation dynamics is dominated by the axial breathing modes; instead, in the trap $V_D$, the axial excitations of the supersolid display a clear coupling to the collective transverse modes. This provides a  signature of the onset of a dimensional crossover, in agreement with the discussion in Ref. \cite{biagioni2022}.

The paper is organized as follows. In Section \ref{sec:system} we introduce the system parameters and the general framework of the extended Gross-Pitaevskii theory for dipolar condensates. Then, in Section \ref{sec:GS} we analyze the equilibrium properties of the condensate in the two trap configurations, $V_C$ and $V_D$, and we characterize the corresponding superfluid-supersolid transition as a function of the $s$-wave scattering length. Section \ref{sec:dynamics} is instead devoted to the dynamical crossing of the transition. We first address, in Sec. \ref{sec:ramp}, the effect of the ramp velocity and then, in Sec. \ref{sec:formation_time} we discuss how the formation time of the supersolid affects the crossing of the transition. Finally, in Sec. \ref{sec:collective_modes} we examine the collective oscillation of the system that arise due to the excess of energy acquired during the ramp across the transition, in both the supersolid and superfluid phases. Concluding remarks are drawn in Section \ref{sec:conclusions}.

\begin{figure*}[t]
     \centerline{\includegraphics[width=0.95\columnwidth]{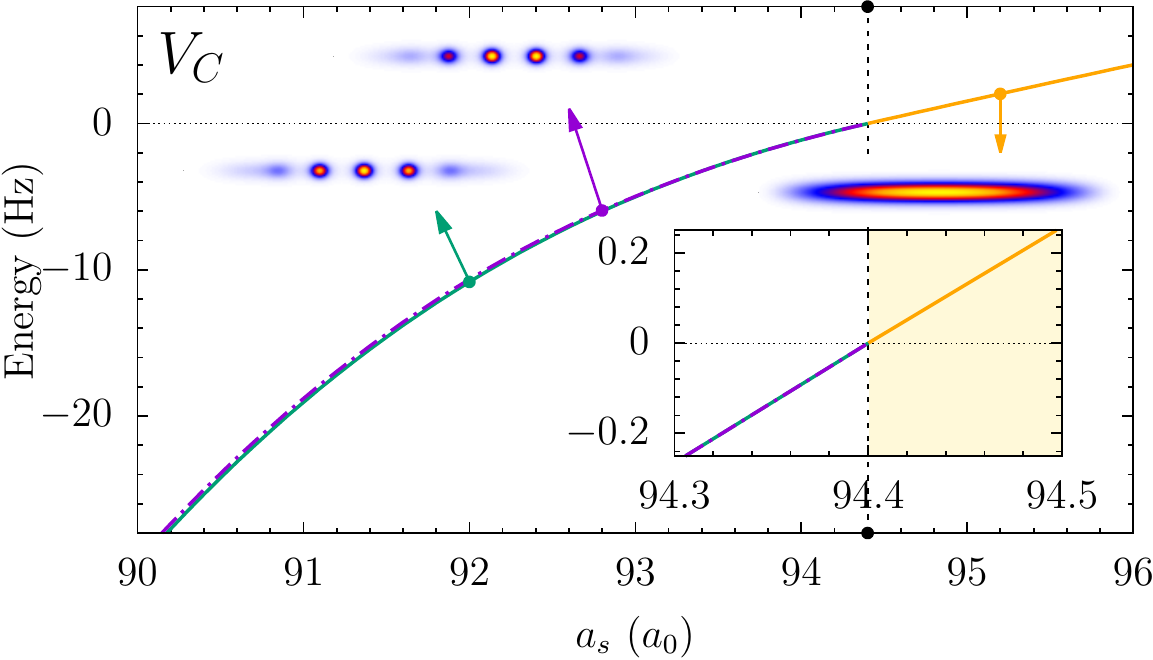}
    \hspace{0.5cm}
    \includegraphics[width=0.95\columnwidth]{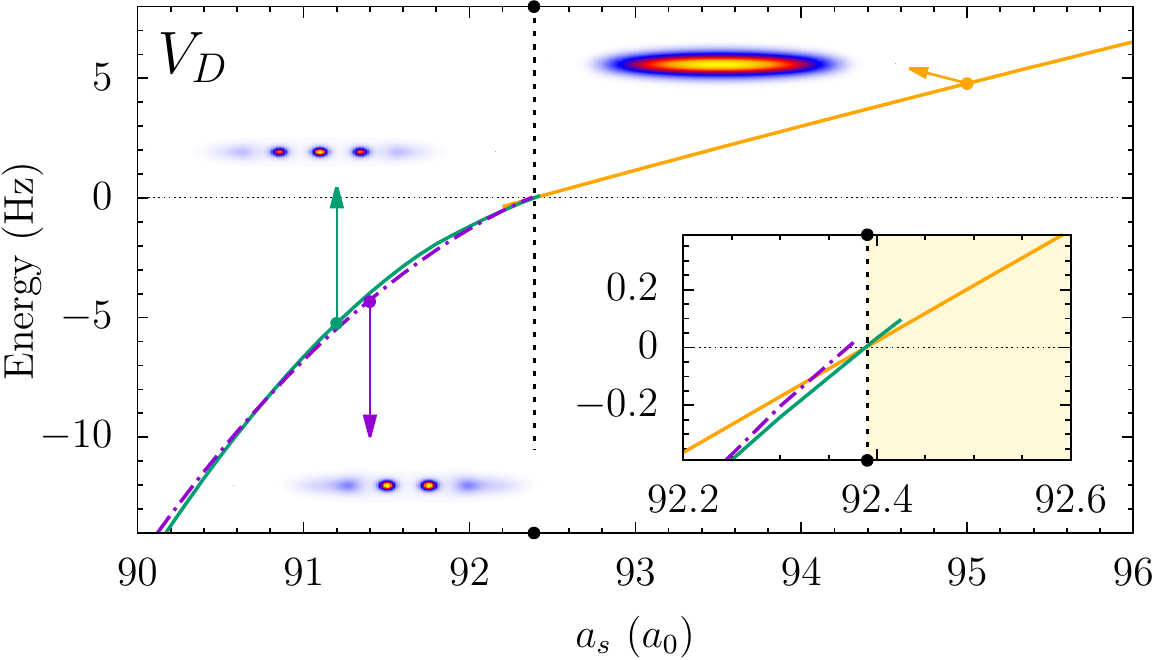}}
    \caption{
    Normalized energy, $(E[a_s] -E[a_s^c])/h$, 
    for the supersolid and superfluid density configurations, as a function of the contact scattering length $a_s$, for the potential $V_{C}$ (left) and $V_{D}$ (right). Typical density distributions (see text; in each plot the color scale is weighted by the density distribution) are indicated by the arrows (the colors of which match the color of the corresponding energy lines). The vertical dashed line represents the boundary between the supersolid and superfluid phases (on its left and right, respectively), at the critical scattering length $a_{s}^{c}$. The insets show the energy behavior in the vicinity of $a_{s}^{c}$.}
    \label{fig:GS}
\end{figure*}

\section{System}
\label{sec:system}

We consider the typical experimental configuration of Ref. \cite{biagioni2022}. A dipolar condensate composed by $N=3\times10^{4}$ magnetic atoms of $^{162}$Dy -- with tunable $s$-wave scattering length $a_{s}$ and dipolar length $a_{dd}=130a_0$ ($a_0$ being the Bohr radius) -- is trapped by a harmonic potential with frequencies $\omega=2\pi\times(\nu_x,\nu_y,\nu_z)$.
As in the experiment, we consider two different trap configurations, namely $\omega_{C}=2\pi\times(15, 101, 94)$ Hz, and $\omega_{D}=2\pi\times(20, 67, 102)$ Hz, where the labels $C/D$ refer to the continuous/discontinuous character of the transition (see Sec. \ref{sec:GS}), in line with the notations employed in Ref. \cite{biagioni2022}. Accordingly, we indicate the corresponding harmonic potentials as $V_{C}$ and $V_{D}$.

This system can be described in terms of a generalized Gross-Pitaevskii (GP) theory including dipolar interactions \cite{Ronen:2006} and the Lee-Huang-Yang correction accounting for quantum fluctuations (within the local density approximation) \cite{Wachtler:2016}. The energy functional can be written as $E = E_{GP} + E_{dd} + E_{LHY}$ with
\begin{align}
E_{GP} &= 
\int \left[\frac{\hbar^2}{2m}|\nabla \psi(\bm{r})|^2  + V_{C/D}(\bm{r})n(\bm{r})+\frac{g}{2} n^2(\bm{r})
\right]d\bm{r}\,,
\nonumber\\
E_{dd} &=\frac{C_{dd}}{2}\iint n(\bm{r})V_{dd}(\bm{r}-\bm{r}')n(\bm{r}') d\bm{r}d\bm{r}'\,,
\label{eq:GPenergy}\\
E_{LHY} &=g_{LHY}\int n^{5/2}(\bm{r})d\bm{r}\,,
\nonumber
\end{align}
where  $E_{GP}=E_{k}+E_{ho}+E_{int}$ is the standard GP energy functional including the kinetic, potential, and contact interaction terms, $V(\bm{r})=(m/2)\sum_{\alpha=x,y,z}\omega_{\alpha}^{2}r_{\alpha}^{2}$ is the harmonic trapping potential, $n(\bm{r})=|\psi(\bm{r})|^2$ represents the condensate density (normalized to the total number of atoms $N$), $g=4\pi\hbar^2 a_{s}/m$ is the contact interaction strength, $V_{dd}(\bm{r})= (1-3\cos^{2}\theta)/(4\pi r^{3})$ the (bare) dipole-dipole potential, $C_{dd}\equiv\mu_{0}\mu^2$ its strength, $\mu$ the modulus of the dipole moment $\bm{\mu}$, $\bm{r}$ the distance between the dipoles, and $\theta$ the angle between the vector $\bm{r}$ and the dipole axis, $\cos\theta=\bm{\mu}\cdot\bm{r}/(\mu r)$. 
As in Ref. \cite{biagioni2022} we consider the magnetic dipoles to be aligned along the $z$ direction by a magnetic field $\bm{B}$.
Finally, the LHY correction is obtained from the expression for homogeneous 3D dipolar condensates under the local-density approximation \cite{Wachtler:2016,Schmitt:2016}. 
The LHY coefficient is $g_{LHY}={256\sqrt{\pi}}{\hbar^{2}a_s^{5/2}}/(15m)\left(1 + 3\epsilon_{dd}^{2}/2\right)$, with $\epsilon_{dd}=\mu_0 \mu^2 N/(3g)$.

\section{Ground State}\label{sec:GS}

We compute the ground state of the system by minimizing numerically the energy functional $E[\psi]$ by means of a conjugate algorithm, see, e.g., Refs. \cite{press2007,Modugno:2003}. In the numerical code the double integral appearing in Eq.\,(\ref{eq:GPenergy}) is mapped into  Fourier space where it can be conveniently computed by means of fast Fourier transform (FFT) algorithms, after regularization \cite{Ronen:2006,politi2022}. 
The behavior of the ground-state energy for the two potentials is shown in Fig. \ref{fig:GS} as a function of the $s$-wave scattering length $a_{s}$, in the range $a_{s}\in[90,96]a_{0}$, along with some representative images of the density distributions in the supersolid (SS) and superfluid (SF) phases \footnote{The full size of the computational box is $54\mu$m$\times18\mu$m$\times18\mu$m. The typical number of grid points in the numerical simulations is $192\times 64 \times 64$}. 
Notice that in the supersolid phase both traps $V_C$ and $V_D$ exhibit two almost degenerate configurations, characterized by either a maximum or a minimum at the center of the trap \footnote{
Numerically, these two states are obtained by using different initial trial wave functions, exploiting the fact that the conjugate gradient algorithm used in the minimization procedure is sensitive to the initial conditions.}. 
These two states are those that -- for symmetry reasons -- survive in the presence of the trap among the infinite equivalent configuration that would be possible in a (infinite) uniform system \footnote{Without an external trapping potential, a change of the lattice phase costs zero energy, therefore in the infinite system the phase of the supersolid lattice is undefined. This is no longer the case in trapped systems, where among all the possible phases, only two minimize the cost in trapping energy. These two states corresponds to the two almost degenerate configurations. characterized by either a maximum or a minimum at the center of the trap.}. 

\begin{figure}[t]
    \centerline{
    \includegraphics[width=\linewidth]{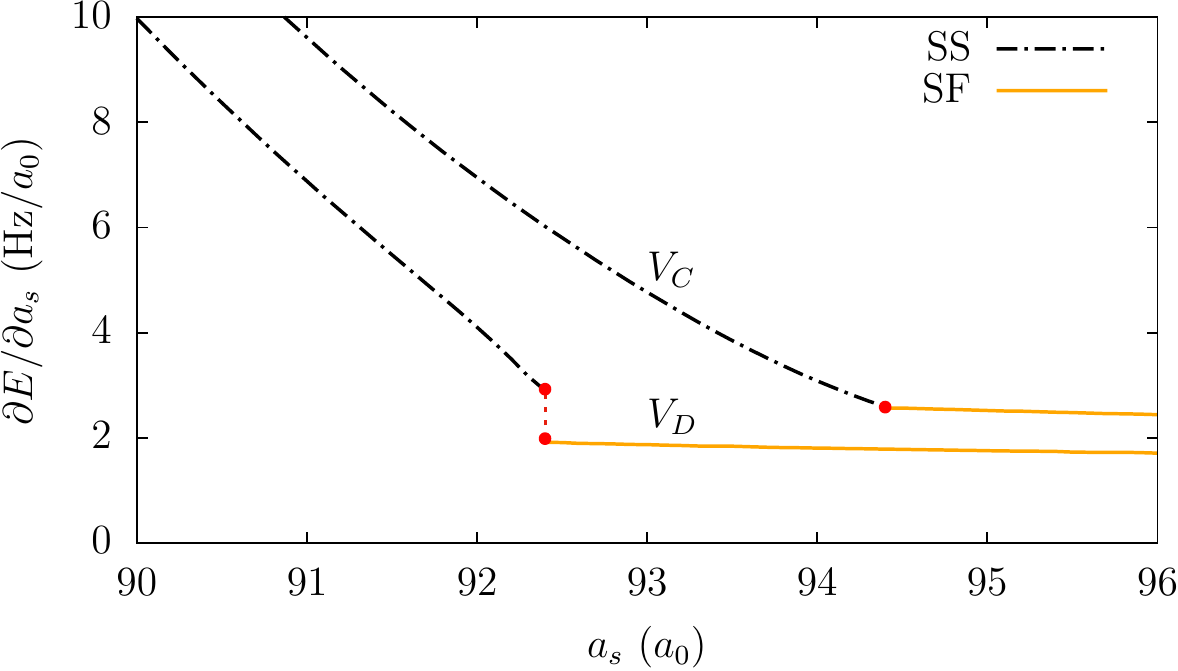}}
    \caption{Derivative of the total energy of the ground state for the two trap configurations, $V_{C}$ and $V_{D}$. The latter exhibits a discontinuity (vertical dashed line) at the transition point (red dots).}
    \label{fig:ener_term}
\end{figure}

Let us start by considering the case of the potential $V_{C}$ (characterized by the tighter confinement along the $y$ direction), shown in the left panel of Fig. \ref{fig:GS}. For this trap, the transition takes place at $a_s^c\simeq 94.4a_0$ and it exhibits a continuous behavior: the superfluid and the supersolid states morph continuously one into the other, as well as their energies [see the inset of Fig. \ref{fig:GS} (left)]. The critical point can be identified, for instance, by the slope change in the first derivative of the energy with respect to the $s$-wave scattering length, see Fig. \ref{fig:ener_term}. The continuity of $\partial E/\partial a_s$ at the critical point confirms the continuous character of the transition in this case (within the numerical precision).

In the case of the potential $V_{D}$, which is characterized by a weaker transverse confinement, the transition takes place at lower value of the scattering length, namely $a_s^c\simeq 92.4a_0$. Remarkably, in this case the two SS states manifest a different behavior in the vicinity of the transition, see the inset of Fig. \ref{fig:GS}b. In particular, the configuration with a maximum at the trap center is the one with lower energy at the boundary with the SF phase, so that $a_s^c$ is actually defined by the crossing of the energy of such a state with that of the SF state. The transition clearly exhibits a discontinuous jump in the first derivative of the energy with respect to the $s$-wave scattering length, see Fig. \ref{fig:ener_term}. As discussed in Ref. \cite{biagioni2022}, this discontinuous behavior of the SF--SS transition is reminiscent of that expected for trapped supersolids with 2D lattice structures \cite{Zh19,Zh21,hertkorn21,Bl21}, and it is due to the fact that even in the case of a single row supersolid the background density may exhibit a triangular structure. 
Even if not visible in the snapshots in Fig. \ref{fig:GS}, this  structure is enhanced when the system is out-of-equilibrium. Indeed, a clear 2D modulation of the background density can be observed during the dynamics of the supersolid discussed in the following section.

\section{Dynamical study of the transition}
\label{sec:dynamics}

\begin{figure}[b]
    \centerline{
    \includegraphics[width=0.9\columnwidth]{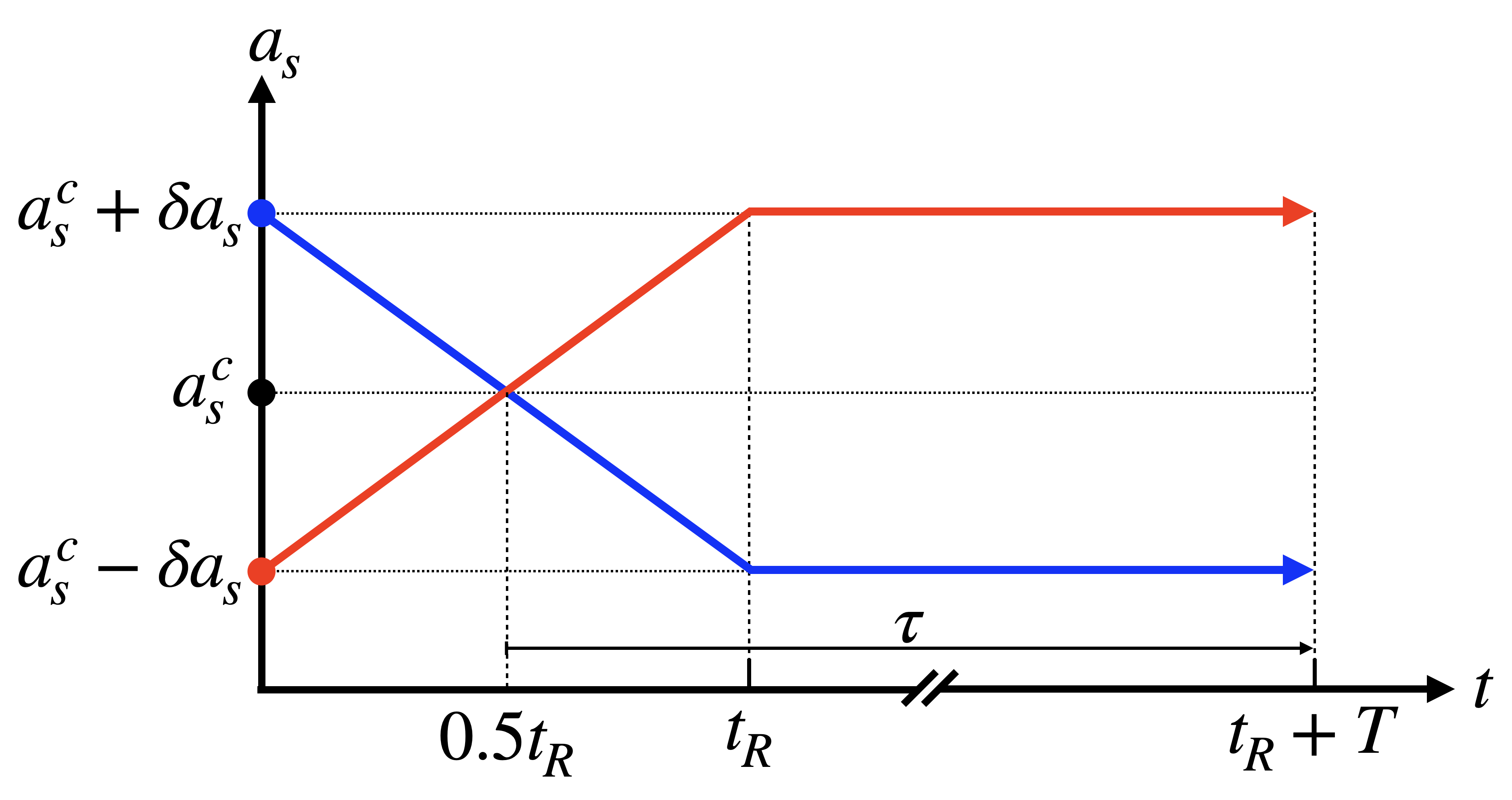}}
    \caption{Scheme of the ramp employed in the numerical simulations: the system is prepared in the ground state either in the superfluid or in the supersolid phase, at $a_{s}^{i}=a_{s}^{c}\pm\delta a_{s}$, and then the scattering length is varied along a linear ramp -- during a time $t_{R}\equiv 2\delta a_{s}/v_{a}$ -- towards a final value in other phase, $a_{s}^{f}=a_{s}^{c}\mp\delta a_{s}$. Then, the system is kept at the final value $a_{s}^{f}$ for a variable time $T$.
   The time spent in the SS/SF phase after crossing the critical value of the scattering length is indicated as $\tau$ (see Sec. \ref{sec:formation_time}).}
    \label{fig:scheme}
\end{figure}

We now turn to the dynamical study of the phase transition, following a protocol  similar to the one employed in the experiment of Ref. \cite{biagioni2022}: the system is initially prepared in a stationary superfluid or supersolid state, at a certain scattering length $a_{s}^{i}$, and then the value of $a_{s}$ is tuned along a linear ramp with constant velocity $da_{s}/dt\equiv v_a$
towards a final value $a_{s}^{f}$, in the other phase. The ramp scheme and the simulation timing are shown in Fig. \ref{fig:scheme}.
For conceptual clarity, here we consider $a_{s}^{i}$ and $a_{s}^{f}$ to be in specular position with respect to the critical point $a_{s}^{c}$, namely $a_{s}^{i/f}=a_{s}^{c}\pm\delta a_{s}$. In the following, we shall consider $\delta a_{s}=1.5a_{0}$ and three ramps with different velocities: i) a ``quench", $v_{a}=\infty$; ii) $v_{a}=0.5$ $a_0/$ms, which corresponds to the nominal velocity employed in the experiment \cite{biagioni2022} (and of the same order of that of Ref. \cite{bottcher2019}); iii) $v_{a}=0.05$ $a_0/$ms, a lower velocity that allows for a quasi adiabatic crossing of the supersolid-superfluid transition in the trap $V_C$ (see Sec. \ref{sec:ramp}), as discussed in Ref. \cite{biagioni2022}. 
This latter choice permits to reproduce a scenario similar to that of the above mentioned experiment, without having to introduce dissipation effects in the theoretical modeling (which are instead present in the experiment \cite{biagioni2022}) \footnote{While those dissipation effects could be easily implemented in the simulations, we refrain from doing so for the sake of clarity and to better identify the underlying dynamics.}.

Therefore, in the following we shall restrict the discussion to the dissipationless scenario, obtained by solving the GP equation \cite{pitaevskii2016}
\begin{equation}\label{eq:gp}
    i\hbar{\partial_{t}\psi}={\delta E[\psi,\psi^*]}/{\delta \psi^*},
\end{equation}
where the energy functional $E[\psi,\psi^*]$ is the one in Eq. (\ref{eq:GPenergy}) \footnote{The GP equation is solved by means of a FFT split-step method, see, e.g.,  Ref. \cite{jackson1998}.}.  
Regarding the two supersolid configurations discussed in the previous section, we notice that during the dynamics across the SF-SS transition the system is likely to select spontaneously the configuration with a maximum at the center of the trap, so that we have used such a configuration also for the initial state of the ramp in the opposite direction, for the sake of simplicity.

\subsection{Effect of the ramp}
\label{sec:ramp}

\begin{figure}[t]
    \centerline{
    \includegraphics[width=0.95\columnwidth]{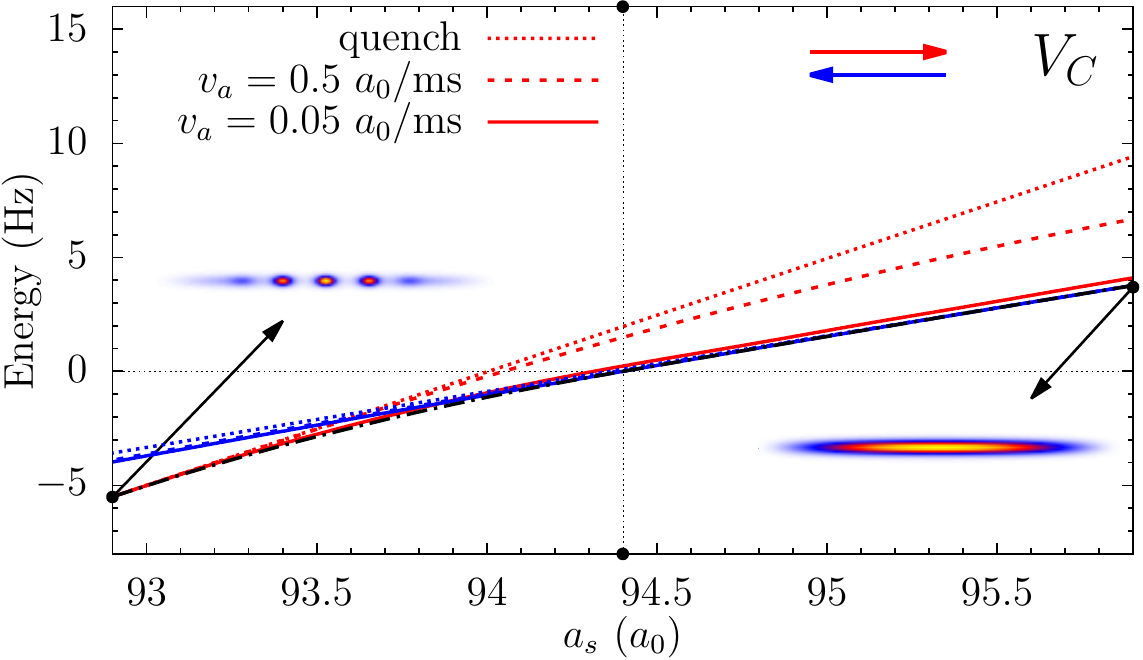}}
    \vspace{0.2cm}
    \centerline{
    \includegraphics[width=0.95\columnwidth]{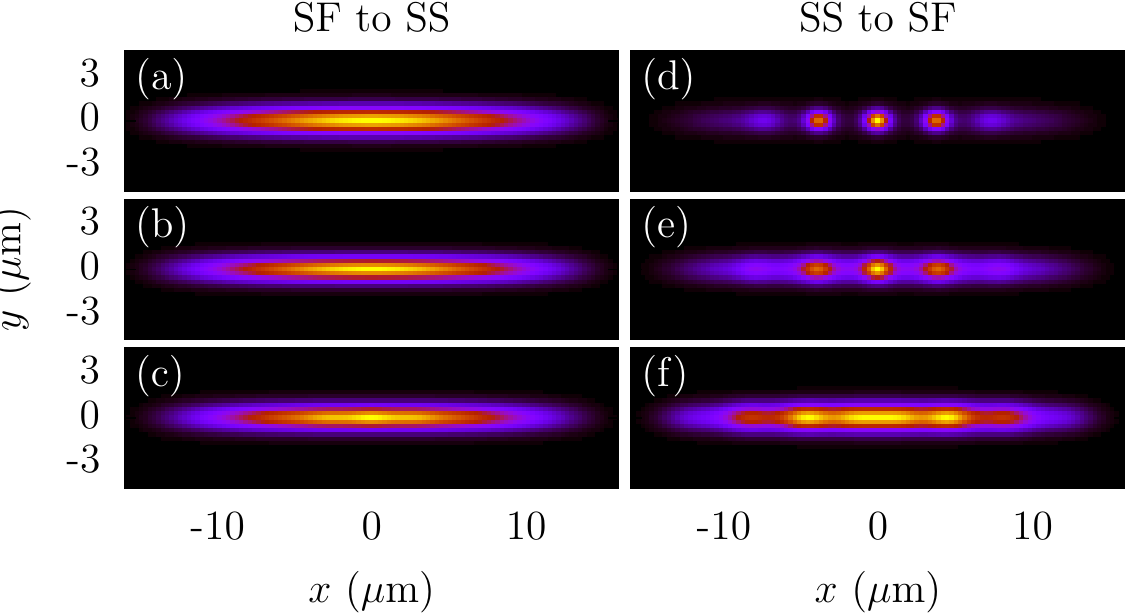}}
    \caption{Behavior of the system along the linear ramp across the SF--SS transition, for the trap $V_C$. (top) Behavior of the 
    normalized energy of the system $(E[a_s] -E_{gs}[a_s^c])/h$, as a function of the scattering length $a_{s}(t)$ during the ramp, for different ramp velocities and directions (see legends).
    The black dots represent the ground state energy. 
    (a-c) Density distribution at the end of the downward ramp from SF to SS, for (a) a quench ($t_{R}=0$ ms), (b) $v_{a}=0.5$ $a_{0}/$ms ($t_{R}=6$ ms), and (c) $v_{a}=0.05$ $a_{0}/$ms ($t_{R}=60$ ms). (d-f) The corresponding distributions after the upward ramp from SS to SF. Notice that the density distributions shown for the case of a quench in panels (a) and (d) correspond, by definition, to the initial density distribution at the beginning of the ramp (also shown as insets in the main panel). In each plot the color scale is weighted by the density distribution.}
    \label{fig4}
\end{figure}
\begin{figure}[t]
    \centerline{
    \includegraphics[width=0.95\columnwidth]{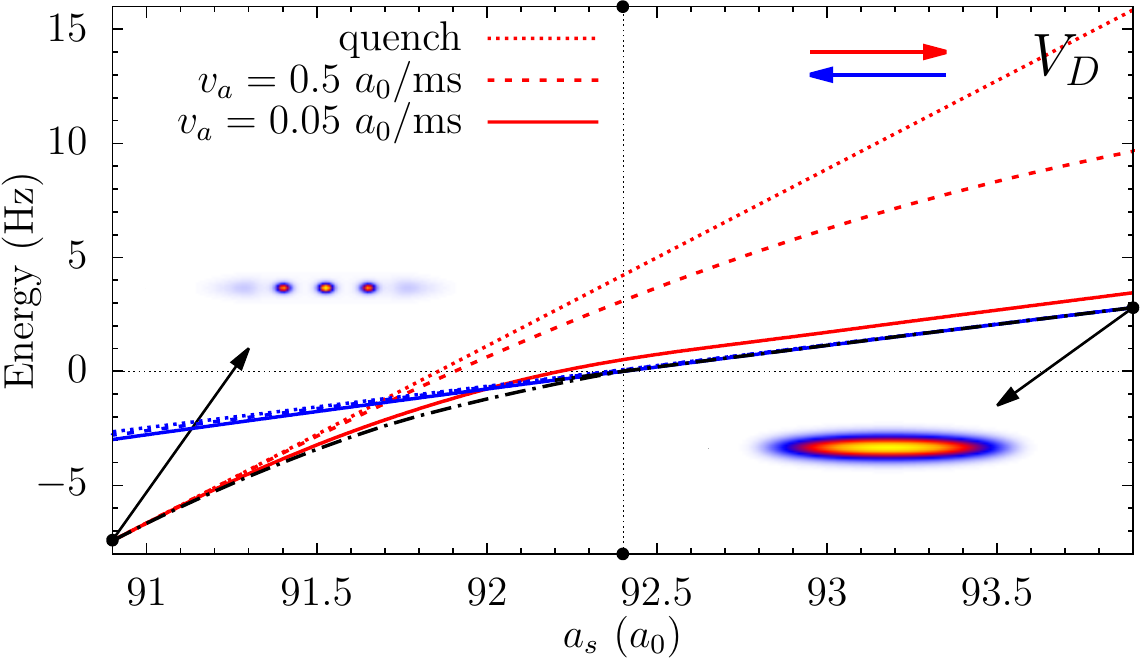}}
    \vspace{0.2cm}
    \centerline{
    \includegraphics[width=0.95\columnwidth]{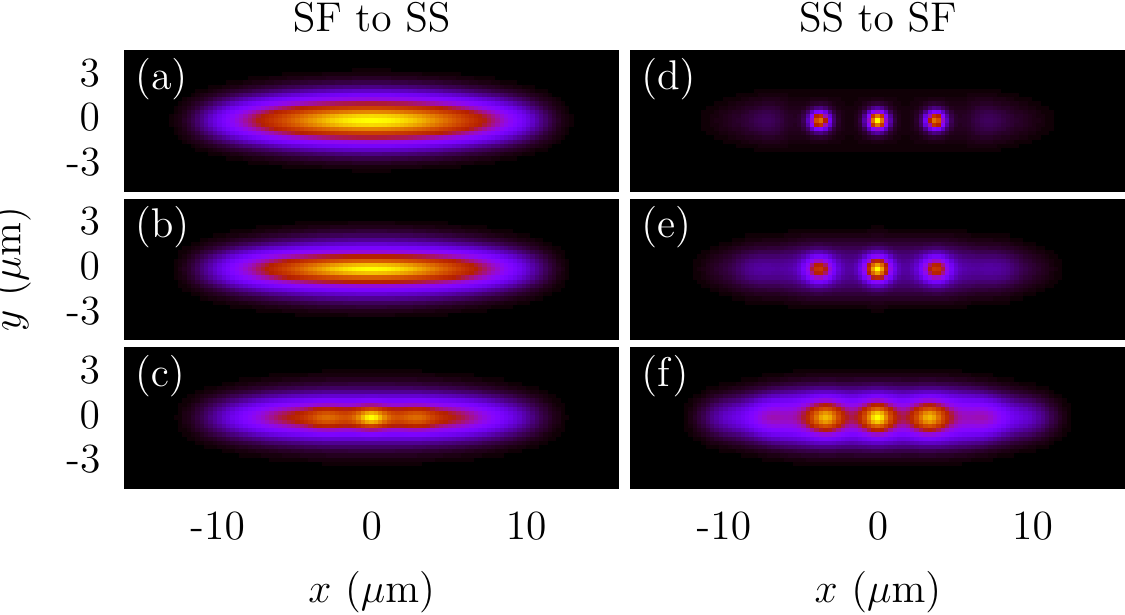}}
    \caption{Behavior of the system along the linear ramp across the SF--SS transition, for the trap $V_D$. The quantities plotted are the same of Fig. \ref{fig4}.}
    \label{fig5}
\end{figure}

Let us now discuss how the system gets modified while varying the scattering length. In particular, we shall first consider how the different ramp velocities affect the energy of the system, and which is the final density distribution of the condensate at the end of each ramp (the dynamics following the end of the ramp will be discussed in Sec. \ref{sec:collective_modes}).  This is shown in Figs. \ref{fig4} and \ref{fig5}, for the traps $V_{C}$ and $V_{D}$, respectively. In the top panel we show the behavior of the energy of the system as a function of the scattering length $a_{s}(t)$ along the three ramps across the SF-SS transition (blue lines) and in the opposite direction, from SS to SF (red lines). 
In the case of the quench, the line is simply a guide to the eye that connects the initial and final value of the scattering length.
The insets represent the initial configurations in the SF and SS phases. 
The density distributions obtained at the end of each ramp, namely at $t=t_R$, are shown in panels (a-c) for the SF-SS transition and in (d-f) for the SS-SF transition.

It is interesting to notice that, when crossing the transition in the downward direction, from SF to SS, both the energy variation and the final density distribution are weakly affected by the ramp velocity.
In the case of a quench, this has to be so because the system is ``projected'' instantaneously in the other quantum phase without changing its density distribution, so that in this case the final configurations 
exactly coincide with the initial ones. 
The case at $v_{a}=0.5$ $a_0/$ms turns out to be almost equivalent to a quench (contrarily to what happens during the SS--SF transition, see below). Only in the case of the slowest ramp at $v_{a}=0.05$ $a_0/$ms in the trap $V_D$ can a slight modulation superimposed to the initial state be appreciated.
The origin of this behavior has to do with the \textit{formation time of the supersolid} (see, e.g., Ref. \cite{bottcher2019}), that will be discussed in Sec. \ref{sec:formation_time}. 

In the opposite direction, when crossing the SS to SF transition, the behavior is quite different: the energy grows linearly if the scattering length is quenched, while it follows the ground state energy almost adiabatically if the scattering length is slowly varied (at $v_{a}=0.05$ $a_{0}/$ms). In addition, it is evident both from the energy behavior and from the final configuration in Fig. \ref{fig4}f that for the trap $V_C$ such a ramp
is sufficiently slow to bring the SS state close to the SF ground-state, with a small excitation energy embedded in a density deformation that is reminiscent of the initial state. 
Remarkably, such a deformation is significantly larger in trap $V_D$ (compare Figs. \ref{fig4}f and \ref{fig5}f). Moreover, the residual energy on the superfluid side is larger in trap $V_D$ than in trap $V_C$ for each of the three ramps (see also the discussion in 
Ref. \cite{biagioni2022}). These observations are consistent with the continuous/discontinuous character of the transition in the two cases.

In order to get further insight on the behaviour of the total energy along the ramp it is convenient to rewrite the energy functional  \eqref{eq:GPenergy} in the following form, which makes the dependence on $a_s$ explicit:
\begin{align}    
\label{eq:fact}
    E[\psi;a_{s}] &= E_{k}[\psi] + E_{ho}[n] +a_{s}{\cal E}^{int}[n] + E_{dd}[n] \\
    \nonumber &\qquad
    + a_s^{5/2}\left[1+\frac{3}{2}\left(\frac{a_{dd}}{a_s}\right)^2\right]{\cal E}^{LHY}[n],
\end{align}
with ${\cal E}^{int}[n]$ and ${\cal E}^{LHY}[n]$  functionals that depend on the density only.
Thus the dependence on the scattering length is explicit for the mean-field interaction energy and the LHY correction, whereas all the other terms depend on $a_s$ only indirectly, through the condensate density  (with the exception of the kinetic term, that is sensitive also to the wave-function phase).

The behavior of the system during the ramp across the transition is therefore characterized by two time scales: the duration of the ramp, $t_{R}=2 \delta a_{s}/v_{a}$, and the timescale 
required for variations in the condensate density to appear.
Remarkably, the latter strongly depends on the phase of the system. In the SS phase, the emergence of the crystalline order of a supersolid is characterized by a minimal formation time, as we shall see in the following section (see also Ref. \cite{bottcher2019}). For the present system, this timescale is of the order of $30$ ms, that corresponds to $t_{R}/2$ for the slowest ramp, at $v_{a}=0.05$ $a_0/$ms.
This implies that during the SF-SS ramp the change in the condensate density is almost negligible, so that the behavior of the energy is dictated only by the explicit dependence on $a_s$. In particular, by considering that $E^{LHY}\!\!\!\!\ll E^{int}$, from Eq. (\ref{eq:fact}) it is evident that the energy in the SS phase can be obtained as a linear continuation of the SF energy branch, as shown in Figs. \ref{fig4} and \ref{fig5}. 
Therefore, the energy variation along the SF-SS ramp
can be written as $\delta E\simeq \delta a_{s}{\cal E}^{int}[n]$.

When going from SS to SF the scenario is quite different because, when the scattering length is increased during the upward ramp, the system can gradually relax the initial supersolid profile into a smoother one, suppressing the crystalline order. Indeed, in the SF phase the relevant time scale is not that of formation, but rather the one characterizing the collective excitations of the condensate.
  This timescale is roughly of the order of $\tau_{0}/4$, where $\tau_{0}$ is the oscillation period of the main excitation mode. In the present case, the dynamics is dominated by the axial breathing mode \cite{Ta19b}, and the associated timescale is of the order of few milliseconds  (see Sec. \ref{sec:formation_time}) \footnote{A similar argument holds also in the SS phase once the supersolid has been formed.}.
This explains the different behavior displayed by the three ramps. Notice that for  the case of the quench the same argument of the SF to SS transition holds, $\delta E\simeq \delta a_{s}{\cal E}^{int}[n]$. More interesting is the behavior that characterizes the ramp at $v_{a}=0.05 a_0/$ms: though the top panels in Figs. \ref{fig4} and \ref{fig5} may suggest that during the upward ramp the system evolves almost adiabatically in both traps $V_{D}$ and $V_{C}$, the fact that the density distributions at the end of the ramp look quite different (see Figs. \ref{fig4}f and \ref{fig5}f) provides instead a signature that the two systems behave differently at the SS-SF transition, with only the one in the trap $V_{C}$ displaying a (quasi) adiabatic behavior.

\subsection{Formation time of the supersolid}
\label{sec:formation_time}

 We now turn to the discussion of the formation time required for the supersolid to emerge after the system has been driven through the SF-SS transition. 
 For convenience, here we introduce the time coordinate $\tau\equiv t-t_{R}/2$, that denotes the time elapsed from the crossing of the SF-SS transition.
 
\begin{figure}[t]
 \centerline{\includegraphics[width=0.9\columnwidth]{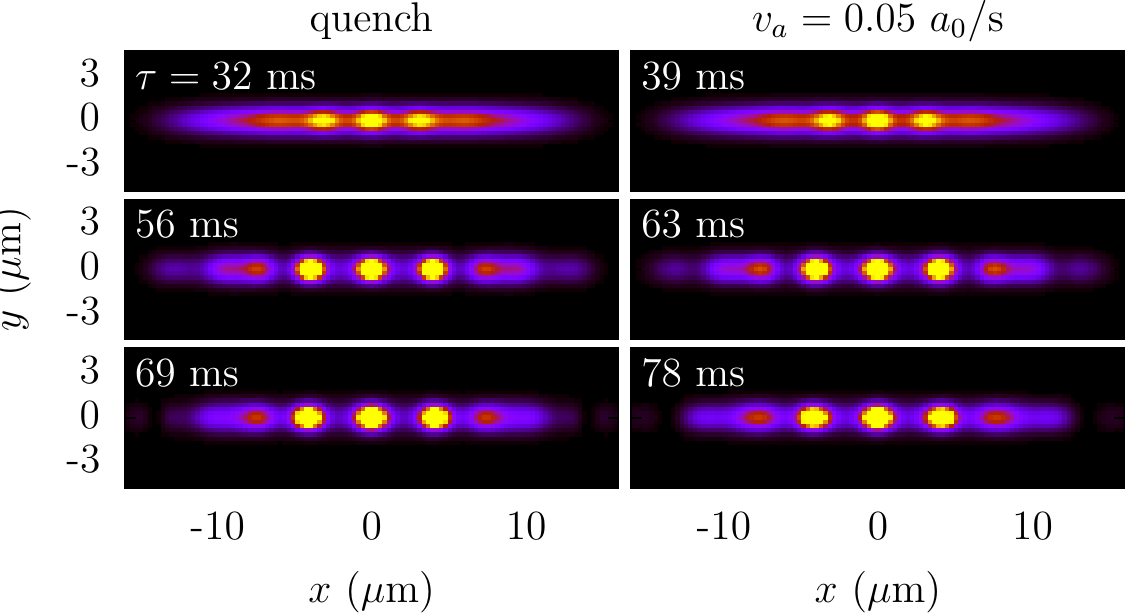}}
    \vspace{0.2cm}
 \centerline{\includegraphics[width=0.9\columnwidth]{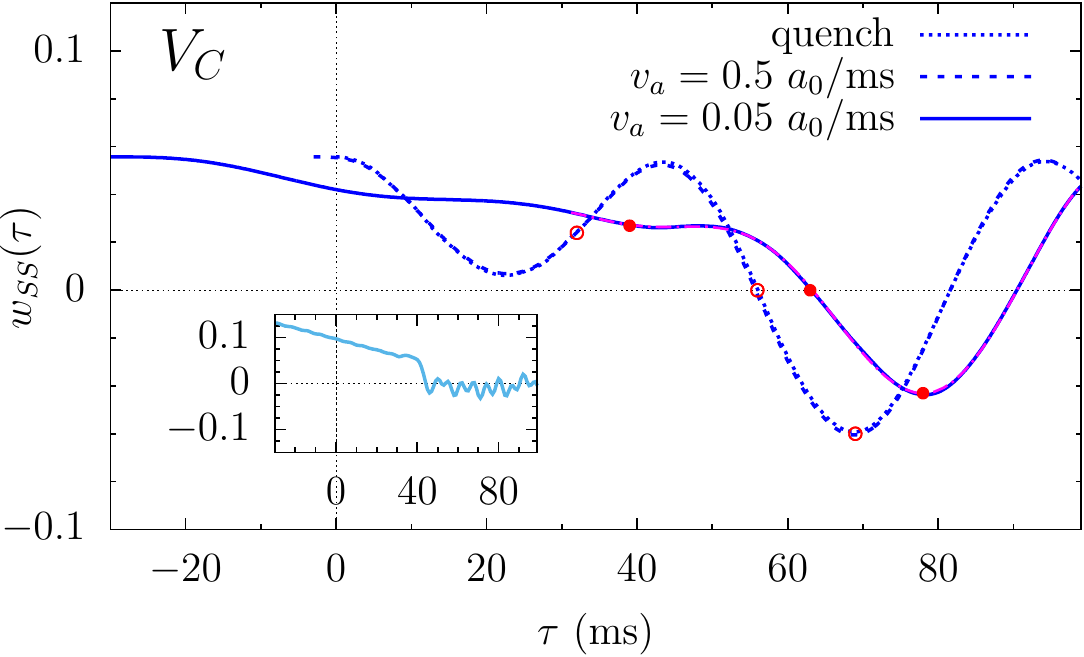}}
 \caption{Dynamical behavior of the system in the SS phase, for the trap $V_C$. 
 The time $\tau=0$ corresponds to the crossing of the SF-SS transition.
 (top)  Density distribution at selected times (see labels) for a quench (left) and for $v_{a}=0.05$ $a_{0}/$ms (right). All densities are saturated at the same level. The case at $v_{a}=0.5$ $a_{0}/$ms, not shown, is very similar to the quench. The panels in the top row correspond to the time at which the supersolid structure starts to emerge clearly, $\tau\approx30\mbox{--}40$ ms. (bottom) Evolution of the axial width $w_{SS}(\tau)$ [see Eq. (\ref{eq:width})], for the three ramp velocities considered here.
The empty/solid red dots corresponds to the density distributions shown in the top panels.
 The (magenta) dot-dashed line represents a sinusoidal fit with two frequencies,  $\nu_{+}\simeq24.0$ (Hz) and $\nu_{-}\simeq13.9$ (Hz), accounting for the doubling of the axial breathing mode of a supersolid (see text). The inset shows the behavior of the transverse width $w_{SS}(\tau)$ along the direction $y$  (for $v_{a}=0.05$ $a_{0}/$ms).}
 \label{fig6}
\end{figure}

\begin{figure}[t]
 \centerline{\includegraphics[width=0.9\columnwidth]{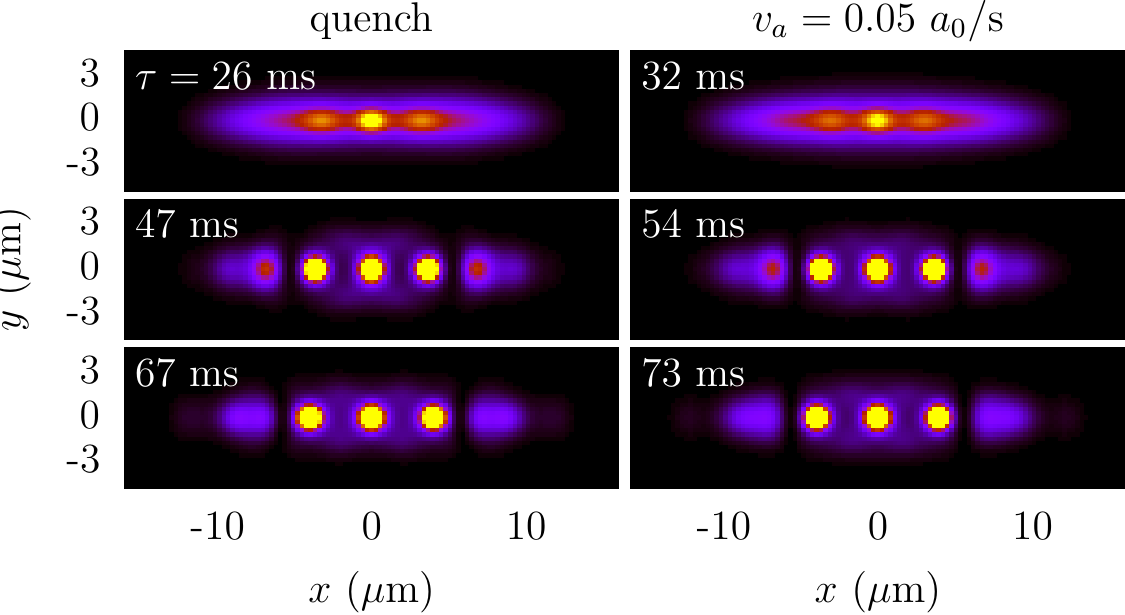}}
\vspace{0.2cm}
 \centerline{\includegraphics[width=0.9\columnwidth]{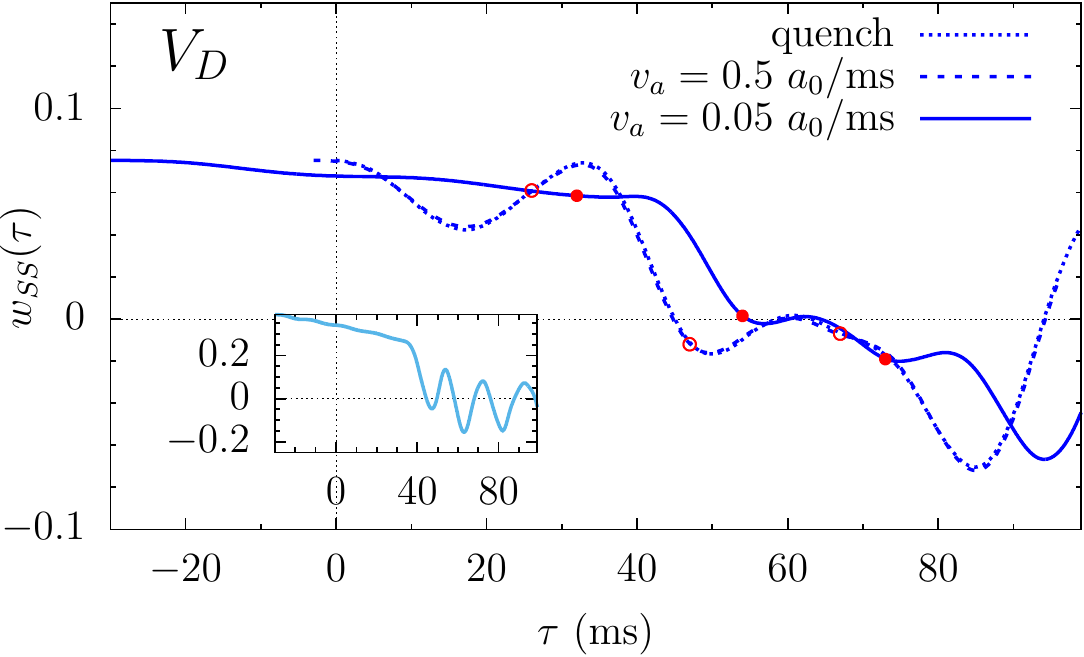}}
 \caption{Dynamical behavior of the system once it has entered the SS phase, for the trap $V_D$. The quantities plotted are the same of Fig. \ref{fig6}. Notice the appearance of a two dimensional structure during the evolution (second and third row of panels), associated to the onset of a transverse oscillation mode shown in the inset (see text). It is also worth noticing that in this trap the supersolid starts to form slightly earlier than as in Fig. \ref{fig6}, at $\tau\approx25\mbox{--}30$ ms (see the top panels).
 }
 \label{fig7}
\end{figure}

In the case of the trap $V_{C}$, we find that the supersolid structure starts to develop at  $\tau_{SS}^{C}\approx 30-40$ ms \footnote{Notice that the formation time observed experimentally in Ref. \cite{biagioni2022} is shorter than the one we find from the present numerical analysis, likely due to finite temperature effects and  three-body losses, see also Refs. \cite{bottcher2019,gallemi2020}.}, see the top row of Fig. \ref{fig6}. Instead, the trap $V_{D}$ is characterized by shorter formation times, $\tau_{SS}^{D}\approx 25-30$ ms, see Fig. \ref{fig7}.
By accident, this formation time is of the order of $t_R/2$ for the ramp at $v_{a}=0.05 a_0/$ms, and this clarifies why in this case the supersolid pattern already emerges at the end of the ramp, see Fig. \ref{fig5}c. 

Remarkably, the figures above also show that $\tau_{SS}$ not only depends on the trap configuration, but also on the ramp velocity. Indeed, as discussed in Ref. \cite{bottcher2019}, longer formation times are associated to a lower energy difference between the initial superfluid configuration (at $\tau=0$) and the target equilibrium supersolid state. 
In the case of a quench, $\tau_{SS}$ depends only on the final energy difference $\Delta E\equiv E_{SF}[a_{s}^{f}] - E_{SS}[a_{s}^{f}]$.
Instead, in the case of a linear ramp the system spends some time in passing from $a_{s}^{c}$ to $a_{s}^{f}$ so that, before reaching the final value of the scattering, it has to cross a region where the energy gap between the SF and SS branches is smaller, see Fig. \ref{fig4}, and this qualitatively explains the slight delay observed. Regarding the difference between the traps $V_{C}$ and $V_{D}$, by comparing Figs. \ref{fig4} and \ref{fig5} it is evident that the 
latter is characterized by a larger gap, which therefore corresponds to a shorter formation time according to the above argument.


\begin{figure*}[t]
    \centerline{\includegraphics[width=0.9\columnwidth]{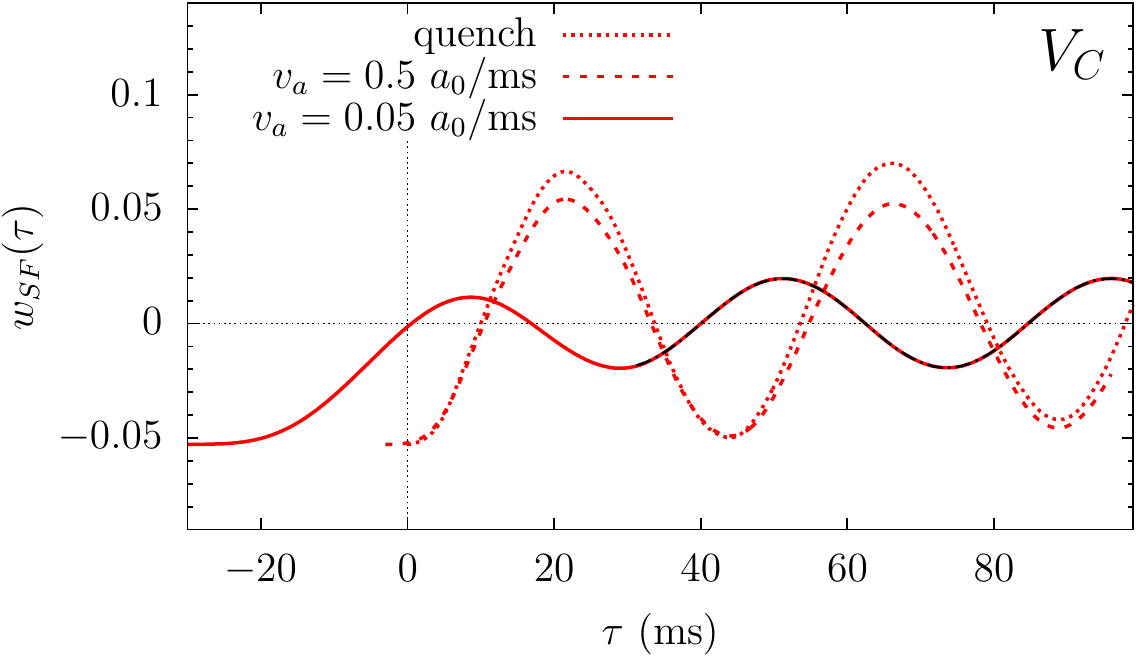}
\hspace{0.15\columnwidth}
\includegraphics[width=0.9\columnwidth]{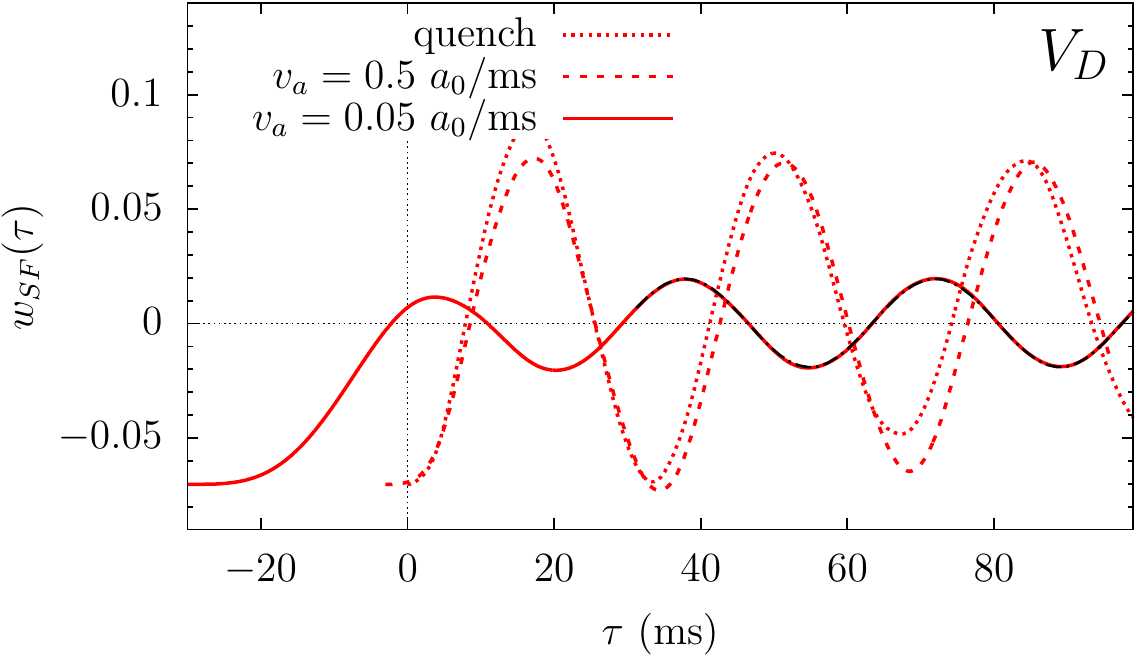}}
\caption{Evolution of the normalized relative width $w_{SF}(\tau)$ defined in Eq. (\ref{eq:width}),
for the potential $V_{C}$ (left) and $V_{D}$ (right). The black dot-dashed line superimposed to the line at $v_{a}=0.05a_0$/ms from the end of the ramp on ($\tau\ge30$ms) represents a sinusoidal fit of the form $A\sin(2\pi\nu t + \theta) + B$. The fit returns $\nu_D\simeq29.3$ (Hz) and $\nu_C\simeq22.7$ (Hz), corresponding to the axial breathing mode (see text).}
\label{fig8}
\end{figure*}

\subsection{Collective oscillations}
\label{sec:collective_modes}

Once the system has entered the new phase, either SS or SF, the excitation energy acquired during the ramp eventually drives the system into collective oscillations. In the present case, since the phase transition mainly affects the density distribution along the axial direction of the trap, the major contribution comes from the so-called axial breathing mode \cite{Ta19b}. The latter can be conveniently characterized by considering the width along the $x$-direction, which corresponds to the axial direction of the supersolid. In order to do so, we define a normalized relative width as
\begin{equation}
     w_{\alpha}(t)\equiv\left[\sigma_{x}(t)-\sigma_{x\alpha}^{eq}\right]/{\sigma_{x\alpha}^{eq}},
    \label{eq:width}
\end{equation}
where $\sigma_{x}^{2}(t)\equiv \langle x^2\rangle=(1/N)\int x^2n(\bm{r},t)d\bm{r}$,
$\alpha$ indicates the SF and SS states, and $\sigma_{x\alpha}^{eq}$ the corresponding equilibrium widths at $a_{s}^{c}\pm\delta a_{s}$, respectively. In the following, we shall consider especially the behavior of the width as a function of the time $\tau$ elapsed from the crossing of the SF-SS transition, namely $w_{\alpha}(\tau)$. This quantity is shown in the bottom panels of Figs. \ref{fig6}, \ref{fig7} for the SS case, and in Fig. \ref{fig8} for the SF case. Overall, the behavior of $w_{\alpha}(\tau)$ provides an additional characterization of the (non)adiabaticity of the various ramps and of the character of the transition for the two trap configurations.

Let us first consider the SS case, in Figs. \ref{fig6}, \ref{fig7}. First of all, we notice that the ramp at $v_{a}=0.5$ $a_{0}/$ms, corresponding to the nominal value employed in the experimental protocol of Ref. \cite{biagioni2022}, is almost indistinguishable from a quench. 
Instead, the slowest ramp at $v_{a}=0.05$ $a_{0}/$ms presents a distinctive feature in the fact that the value of the width decreases gradually along the ramp, $-30$ ms $<\tau<30$ ms, indicating that the system is able to smoothly adjust its shape to the changing value of the scattering length, from $a_{s}^{i}$ to $a_{s}^{f}$. 

As discussed in  Ref. \cite{Ta19b}, the excitation dynamics of an elongated supersolid is characterized by a \textit{doubling} of the axial breathing mode of a dipolar condensate (in the SF regime, see later on). The two modes that appear in the SS phase are associated to the deformation of the supersolid lattice structure, namely its amplitude and spacing. The former is dominated by the higher-frequency mode, the latter by the lower-frequency one. 
In the case of the trap $V_C$ we observe indeed a beating of two frequencies, see Fig. \ref{fig6}, that we fit with a sinusoidal function of the form $A_{+}\sin(2\pi\nu_{+} t + \theta_{+}) + A_{-}\sin(2\pi\nu_{-} t + \theta_{-}) + B$ (see the magenta dot-dashed lines in the figure). The fit of the curve at $v_{a}=0.05$ $a_{0}/$ms returns $(\nu_{+}/\nu_x)_C\simeq1.6$ and $(\nu_{-}/\nu_x)_C\simeq0.93$, which is consistent with the picture provided in  Ref. \cite{Ta19b} (here $\epsilon_{dd}^C=1.39$)  \footnote{A direct quantitative comparison with Ref. \cite{Ta19b} is not possible because the actual values of the frequencies depend on the parameters of the system, which are different.}.
A similar result is obtained from the fit of the other two lines. 
In the transverse directions we not see instead any significant oscillation, as one can see from the behavior of transverse width along the $y$ direction shown in the inset.
In the figure we also show some snapshot of the density distribution, at selected
times: when the supersolid structure starts to emerge clearly (top row, discussed previously), at  $w_{SS}(\tau)=0$ (middle row), and when $w_{SS}(\tau)$ first reaches a minimum of the oscillation (bottom row). These snapshots well represent the qualitative behavior along the whole dynamics considered here (also when $w_{SS}(\tau)$ gets to an oscillation maximum), which can be indeed fully characterized by the deformation of the supersolid structure discussed previously, affecting the amplitude and the spacing along the axial direction.

Instead, the case of the trap $V_D$ presents a distinctive behavior associated with the emergence of a characteristic pattern in the background density distribution. This is visible in the top panels of Fig. \ref{fig7} (middle and bottom rows). Remarkably, this pattern is reminiscent of the triangular lattice structure expected for 2D supersolids \cite{Zh19,Zh21,hertkorn21,Bl21}, see also the discussion in Ref. \cite{biagioni2022}. We find that the corresponding transverse excitation mode (shown in the inset) is characterized by a relatively high frequency, $\nu_{\perp}\simeq60$Hz, which couples with the axial breathing modes. This accounts for the ``fast'' oscillations that are visible in the continuous line at $v_{a}=0.05$ $a_{0}/$ms in Fig. \ref{fig7}. As a matter of fact, a clean sinusoidal fit (with two or even three frequencies) is not possible in this case.

Finally, let us consider the excitation produced by the ramps in the opposite direction, when the system is driven into the SF phase. In this case we find that the condensate oscillations are dominated by a single excitation mode, namely the axial breathing mode of a dipolar condensate discussed previously, see Fig. \ref{fig8}. This holds for both trap configurations, $V_C$ and $V_D$. The corresponding frequency is expected to be slightly below the mean-field solution for the breathing mode frequency of a superfluid without dipolar interactions, $\omega=\sqrt{5/2}\omega_{x}$ \cite{stringari1996}, see again Ref. \cite{Ta19b}. In particular, we find $(\nu/\nu_x)_C\simeq1.51$ ($\epsilon_{dd}^C=1.35$) and $(\nu/\nu_x)_D\simeq1.47$ ($\epsilon_{dd}^D=1.37$).

\section{Conclusions}
\label{sec:conclusions}

We have presented a theoretical discussion -- within the framework of the extended Gross-Pitaevskii theory including Lee-Huang-Yang quantum corrections -- of the superfluid-supersolid transition of an elongated dipolar condensate as reported in the recent experiment by G. Biagioni \textit{et al.} [Phys. Rev. X \textbf{12}, 021019 (2022)]. We have considered both trapping potentials employed in the experiment, providing a characterization of the equilibrium and dynamical properties of the system as a function of the inter-particle scattering length, which is the parameter that is varied experimentally for driving the transition. Although both traps display a one row supersolid (for $a_s < a_s^c$), already at the level of the ground state the two traps present a distinctive behavior. For a sufficiently strong transverse confinement (the trap $V_C$) the SF-SS transition has a smooth continuous character, as it is expected for (quasi) 1D systems, with the superfluid and the supersolid states morphing continuously one into the other, as well as their energies. Instead, in the case of the potential $V_{D}$, which is characterized by a weaker transverse confinement, the transition clearly exhibits a discontinuous jump in the first derivative of the energy with respect to the $s$-wave scattering length, as it is expected for trapped supersolids with 2D lattice structures \cite{Zh19,Zh21,hertkorn21,Bl21}. These properties reflects in the collective oscillations of the system, when the scattering length is dynamically ramped across the transition, from one phase to the other. In particular, we find that when the system is driven quasi adiabatically into the superfluid phase the system performs clean axial breathing oscillations, in both traps. In the opposite direction, the situation is quite different: in the trap $V_C$ the excitation dynamics is still dominated by the doubling of the axial breathing modes, whereas when the transition becomes discontinuous, in the trap $V_D$, the collective excitations of the supersolid display a coupling with the transverse modes, signalling the onset of a dimensional crossover. These findings provide further insights on the experimental results reported in Ref. \cite{biagioni2022}.

\begin{acknowledgments}
We thank G. Modugno, L. Tanzi, C. Gabbanini and A. Fioretti from the Pisa-Florence Dysprosium lab for support and useful discussions.
This work was supported by Grant PGC2018-101355-B-I00 funded by MCIN/AEI/10.13039/501100011033, by “ERDF A way of making Europe”, by the Basque Government through Grant No. IT1470-22, and by QuantERA grant MAQS, by CNR-INO.
\end{acknowledgments}


%

\end{document}